\documentclass{article}
\usepackage{PRIMEarxiv}

\usepackage[utf8]{inputenc} 
\usepackage[T1]{fontenc}    
\usepackage{hyperref}       
\usepackage{url}            
\usepackage{booktabs}       
\usepackage{amsfonts}       
\usepackage{nicefrac}       
\usepackage{microtype}      
\usepackage{lipsum}
\usepackage{fancyhdr}       
\usepackage{graphicx}       
\graphicspath{{media/}}     

\usepackage{amsmath}
\usepackage{stmaryrd}
\usepackage{amssymb}

\usepackage{enumitem} 

\usepackage{tikz}
\usetikzlibrary{shapes.geometric, arrows, positioning, calc, arrows.meta, backgrounds, fit}

\usepackage[most]{tcolorbox}

\title{Correct-by-Construction G-Code Generation: A Neuro-Symbolic Approach via Separation Logic
}

\author{
  Yeonseok Lee \\
  SLING AI Inc. \\
  Incheon, Republic of Korea \\
  \texttt{ylee@sling.ai.kr}
}

\begin{document}

\newcommand{\bigast}{\scalebox{1.5}{$\ast$}}

\maketitle

\begin{abstract}
This paper proposes a neuro-symbolic framework for G-code generation that seeks to integrate the neural generative capabilities of the GLLM method (Abdelaal et al., 2025) with formal verification via a Separation Logic (SL) prover. To establish a reliable physical baseline, the framework extracts deterministic boundary representations from 3D CAD models (STEP files) using the OpenCASCADE framework. This extracted geometric data supports a two-component architecture: the LLM serves as an initial code generator, while the SL Prover, utilizing a Spatial Heap model, evaluates the output. By conceptualizing physical collisions as logical Spatial Data Races---violations of the separating conjunction in SL---our framework translates proof failures into structured mathematical feedback. These failures are condensed into bounding boxes that serve as directives for the LLM's iterative self-correction. Ultimately, this work aims to develop a self-correcting system that reduces the need for human supervision, leading to safer and verified autonomous manufacturing.
\end{abstract}

\keywords{Neuro-Symbolic AI \and Large Language Models \and Separation Logic \and G-code \and Formal Verification \and Collision Avoidance \and OpenCASCADE}

\section{Introduction}

\subsection{The Challenge of G-Code Generation and LLMs}
The domain of Computer Numerical Control (CNC) machining stands at a critical juncture. Large Language Models (LLMs) offer unprecedented potential to automate low-level machine instruction synthesis (e.g., G-code) via natural language commands. Systems such as GLLM \cite{abdelaal2025gllm} have successfully utilized fine-tuned models, structured prompts, and Retrieval-Augmented Generation (RAG) to translate human intent into executable code. 
However, deploying these generative models directly into cyber-physical systems introduces unique challenges regarding absolute physical safety. Even when an LLM successfully generates G-code that is syntactically perfect and semantically aligned with the target geometry \cite{abdelaal2025gllm}, safe execution remains highly dependent on the specific, dynamic state of the machine and its environment. Variables such as updated workholding fixtures, adjustments in stock material dimensions, or rapid changes in manufacturing requirements mean that a logically sound toolpath may still result in an unintended physical collision. Furthermore, the generated code must eventually account for mechanical realities such as servo lag or spindle runout. Consequently, generating accurate G-code is the first step; an independent, deterministic verification mechanism is  required after the generation phase to mathematically guarantee spatial disjointness and safe execution within the physical workspace.

\subsection{The Shift to Formal Verification}
Current industrial CNC safety mechanisms rely on geometric simulations to catch these errors. While simulation is effective for visualizing specific, predefined paths, it lacks symbolic, mathematical proofs of safety and remains vulnerable to corner-case collisions driven by the continuous nature of physical movement. Traditional Computer Aided Manufacturing (CAM) tools calculate worst-case safety margins through testing.

Recent research has pivoted toward formal verification, specifically conceptualizing the continuous physical CNC workspace as a discrete ``Spatial Heap'' \cite{lee2026cnc}. 
By treating physical occupancy as a managed logical resource, this approach allows for the application of Separation Logic (SL) \cite{reynolds2002separation, o2004resources} to formally verify toolpaths, mathematically redefining physical collisions as logical contradictions.

\subsection{The Proposed Neuro-Symbolic Integration}
To resolve the tension between generative creativity and deterministic physical constraints, this paper proposes a neuro-symbolic framework. 
We cast the LLM \cite{abdelaal2025gllm} as a \textit{Generator} agent and introduce a domain-specific Separation Logic (SL) Prover \cite{lee2026cnc} as a deterministic \textit{Verifier}. 
Our methodology establishes a pipeline that strictly decouples continuous physical kinematics from logical spatial evaluation, bypassing dynamic geometric calculation during the proof phase. 
Instead, the pipeline begins by extracting deterministic Boundary Representations (B-Rep) directly from standard 3D CAD (STEP) files using the OpenCASCADE framework \cite{opencascade}. A Parser then utilizes a Zero-Store Model to statically evaluate and exhaust kinematic variables based on this extracted topography. It applies a discretization function alongside Minkowski sum dilations \cite{lozano1983spatial, abrams2000computing} to encapsulate physical uncertainty. 
By the time instructions reach the SL Prover, the environment is represented entirely as a finite set of discrete spatial literals, permitting decidable logical verification.

\subsection{Key Contributions}
Building upon the generative advancements established by the GLLM framework \cite{abdelaal2025gllm} and the formal spatial reasoning of the Spatial Heap \cite{lee2026cnc}, this research proposes a synergistic neuro-symbolic approach aimed at establishing a formally verified, correct-by-construction methodology for AI-driven manufacturing.
Our key contributions focus on the end-to-end integration of raw geometry, neural generation, and formal logic:
\begin{itemize}
    \item \textbf{Deterministic Physical Grounding from STEP files:} Before any generation occurs, the framework addresses the LLM hallucination problem by extracting explicit mathematical parameters (B-Rep) directly from standard STEP files. This continuous spatial data establishes an  physical ground truth, which is mathematically padded and injected into the LLM via Retrieval-Augmented Generation (RAG) to constrain initial toolpath synthesis.

    \item \textbf{A Generator-Verifier Neuro-Symbolic Architecture:} We introduce a  two-way framework that bridges probabilistic AI generation with deterministic formal verification. The Large Language Model operates as a  ``Generator'' that translates informal manufacturing intent into preliminary toolpaths. Simultaneously, our SL Prover \cite{lee2026cnc} acts as a  ``Verifier'', utilizing Separation Logic \cite{reynolds2002separation,berdine2005symbolic} to evaluate the neural output.
    
    \item \textbf{Symbolic-to-Neural Translation via Spatial Data Races:} We establish a formal mechanism for the symbolic engine to communicate with and constrain the neural engine. When the SL Prover detects a violation of the Separating Conjunction ($*$)---indicating that the machine tool's swept volume intersects with restricted environmental space---it triggers a \textit{Spatial Data Race} \cite{lee2026cnc}. Our framework mathematically localizes this logical contradiction and translates the spatial violation into a structured, machine-readable directive.
    
    \item \textbf{Deterministic, Automated Self-Correction:} While contemporary generative approaches have demonstrated the value of self-correction using empirical metrics \cite{abdelaal2025gllm}, our framework introduces a formal alternative. The SL Prover condenses conflicting voxels from a Spatial Data Race into a minimal, precise bounding box ($B_{conflict}$). This deterministic mathematical feedback provides structured spatial bounds to the LLM, autonomously guiding the neural model to refine its generation until a formal proof of safety is achieved.
\end{itemize}

\section{Background and Related Work}

\subsection{LLM-based G-code Generation and GLLM}
Recent advancements in Large Language Models (LLMs) have significantly impacted automated code synthesis. While general-purpose models demonstrate high proficiency in standard programming languages, the synthesis of machine-specific G-code requires deeper domain grounding. Drawing inspiration from these advancements, the GLLM framework \cite{abdelaal2025gllm} introduced a specialized approach for Computer Numerical Control (CNC) machining. 

GLLM utilizes a fine-tuned StarCoder-3B model, optimized through domain-specific training data to bridge the gap between natural language task descriptions and executable RS-274 commands. A core innovation of GLLM is its integrated Retrieval-Augmented Generation (RAG) mechanism, which allows the model to adapt to the high variability of G-code dialects. Furthermore, GLLM employs a robust \textit{Generator-Evaluator} loop for self-correction, utilizing empirical metrics like the Hausdorff distance to measure path similarity against safe trajectories. This pioneering work provides a strong foundation for practical G-code generation; our research seeks to build upon this by introducing a formal mathematical layer that ensures safety.

\subsection{Neuro-Symbolic Foundations and Intelligent Formal Methods}
To address the inherent limitations of purely neural generation—such as the lack of deterministic safety guarantees—the field is increasingly shifting toward neuro-symbolic architectures \cite{neider2022intelligent}. These methods combine the inductive, creative capabilities of neural networks with the deductive, deterministic rigor of formal logic, advancing the vision of ``intelligent formal methods'' \cite{neider2022intelligent}. 

A prominent contemporary example of this paradigm is Neural Model Checking \cite{giacobbe2024neural}, which utilizes neural networks to generate formal proof certificates for Linear Temporal Logic (LTL) specifications. Our research extends this neuro-symbolic paradigm into the physical and spatial domain. Rather than using the prover as a passive safety net, we utilize it as an active, deterministic guide. By identifying specific logical contradictions in the toolpath, the symbolic engine provides structured feedback that constrains the LLM's generative process, ensuring the output is correct-by-construction.

\subsection{Separation Logic and Symbolic Execution}
Separation Logic (SL), pioneered by Reynolds and O'Hearn \cite{reynolds2002separation, o2004resources}, extended Hoare logic \cite{hoare1969axiomatic} to reason about programs with shared mutable state. The defining feature of SL is the \textit{separating conjunction} ($P * Q$), which asserts that the heap can be partitioned into two disjoint portions. This facilitates \textit{local reasoning} via the \textit{Frame Rule}:
\[
\frac{\{P\} \mathbb{C} \{Q\}}{\{P * R\} \mathbb{C} \{Q * R\}}
\]
The Frame Rule ensures that if a command $\mathbb{C}$ safely modifies a part of the state $P$, it will not affect the disjoint ``frame'' $R$. This property is critical for the scalability of verifying complex CNC toolpaths.

Further developments by Berdine et al. \cite{berdine2005symbolic} introduced \textit{Symbolic Execution} within Separation Logic. This technique allows a prover to represent the program state as a ``symbolic heap'' and execute instructions abstractly. Each instruction updates the heap according to its operational semantics. 
This methodology is suited for G-code, where movement commands (e.g., G01) can be modeled as state transitions that ``consume'' and ``produce'' spatial resources.

\subsection{Separation Logic for CNC}
Building upon the SL with pointer arithmetic \cite{brotherston2018complexity} and symbolic execution \cite{berdine2005symbolic}, 
Lee \cite{lee2026cnc} adapts the concept of SL to the physical manufacturing domain. 
In this paradigm, physical occupancy is treated as a managed logical resource. The traditional memory heap is replaced by a coordinate of discrete spatial addresses $c \in \mathbb{Z}^3$, where each address maps to a physical state (e.g., Tool, Environment, Empty, or Stock).

As established in \cite{lee2026cnc}, verifying a G-code program involves a \textit{Parser-Prover Handshake}. A parser exhausts continuous kinematics into discrete spatial literals, which the SL Prover then evaluates. 
Physical collisions are redefined as \textit{Spatial Data Races}, detected when the separating conjunction fails to establish disjointness:
\[
(0,1,0) \mapsto \text{Tool} * (0,1,0) \mapsto \text{Environment}
\]
If the coordinate mapping \text{Tool} intersects with the coordinate mapping \text{Environment}, the proof fails. 
Our framework leverages this failure to generate precise bounding boxes ($B_{conflict}$), which are fed back to the LLM as part of the neuro-symbolic corrective loop.

\section{The Neuro-Symbolic Architecture}

\subsection{The Pipeline}
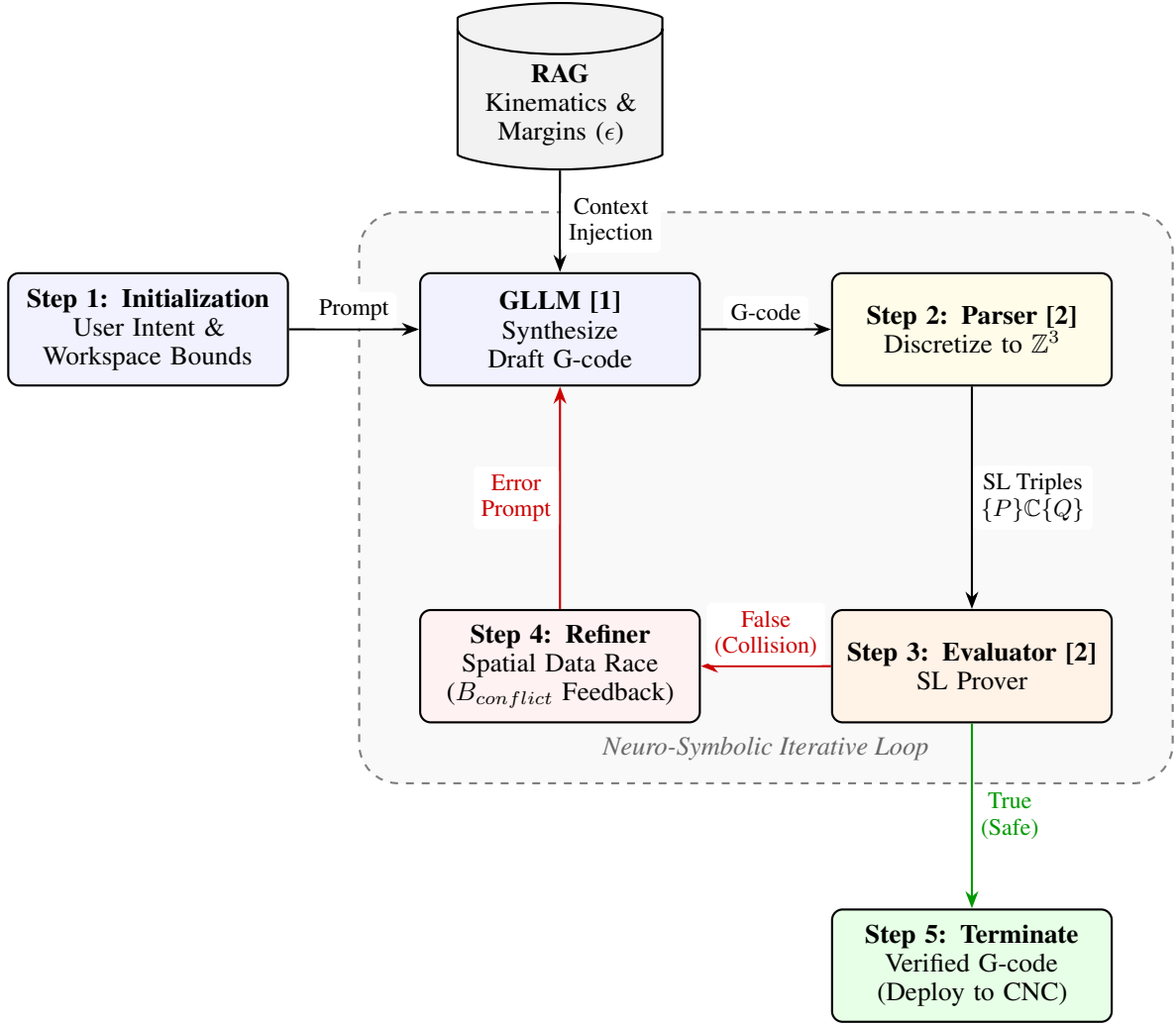
\begin{figure}[htbp]
\centering
\begin{tikzpicture}[
    base/.style={draw=black, thick, align=center, minimum height=1.5cm, rounded corners},
    generator/.style={base, rectangle, fill=blue!5, text width=3.5cm},
    parser/.style={base, rectangle, fill=yellow!10, text width=3.5cm},
    evaluator/.style={base, rectangle, fill=orange!10, text width=3.5cm},
    feedback/.style={base, rectangle, fill=red!5, text width=3.5cm},
    success/.style={base, rectangle, fill=green!10, text width=3.5cm},
    rag/.style={base, cylinder, shape border rotate=90, aspect=0.25, fill=gray!10, text width=2.5cm, minimum height=1.5cm},
    arrow/.style={thick, ->, >=Stealth},
    arrow_text/.style={fill=white, font=\footnotesize, align=center, inner sep=3pt, rounded corners=2pt}
]

\node[generator] (user) at (0, 0) {\textbf{Step 1: Initialization} \\ User Intent \& \\ Workspace Bounds};
\node[generator] (llm) at (5.5, 0) {\textbf{GLLM \cite{abdelaal2025gllm}} \\ Synthesize Draft G-code};
\node[rag] (rag_db) at (5.5, 3) {\textbf{RAG} \\ Kinematics \& \\ Margins ($\epsilon$)};

\node[parser] (parser) at (11, 0) {\textbf{Step 2: Parser \cite{lee2026cnc}} \\ Discretize to $\mathbb{Z}^3$};
\node[evaluator] (prover) at (11, -4.5) {\textbf{Step 3: Evaluator \cite{lee2026cnc}} \\ SL Prover};

\node[feedback] (error) at (5.5, -4.5) {\textbf{Step 4: Refiner} \\ Spatial Data Race \\ ($B_{conflict}$ Feedback)};
\node[success] (deploy) at (11, -8.5) {\textbf{Step 5: Terminate} \\ Verified G-code \\ (Deploy to CNC)};

\draw[arrow] (user) -- node[above, arrow_text] {Prompt} (llm);
\draw[arrow] (rag_db) -- node[right, arrow_text] {Context \\ Injection} (llm);
\draw[arrow] (llm) -- node[above, arrow_text] {G-code} (parser);
\draw[arrow] (parser) -- node[right, arrow_text] {SL Triples \\ $\{P\}\mathbb{C}\{Q\}$} (prover);

\draw[arrow, color=red!80!black, thick] (prover) -- node[above, arrow_text] {False \\ (Collision)} (error);
\draw[arrow, color=red!80!black, thick] (error) -- node[left, arrow_text] {Error \\ Prompt} (llm);

\draw[arrow, color=green!60!black, thick] (prover) -- node[right, arrow_text] {True \\ (Safe)} (deploy);

\begin{scope}[on background layer]
    \node[draw=gray, dashed, thick, rounded corners=15pt, fit=(llm) (parser) (prover) (error), inner sep=0.8cm, fill=gray!5] (loop_box) {};
    \node[anchor=south, font=\itshape\color{gray!80!black}, yshift=0.2cm] at (loop_box.south) {Neuro-Symbolic Iterative Loop};
\end{scope}

\end{tikzpicture}
\caption{The Evaluator-Refiner Pipeline. The Large Language Model generates candidate toolpaths, which are discretized and passed to the Separation Logic (SL) prover. If the SL evaluation fails to prove spatial disjointness ($*$), a Spatial Data Race is triggered, sending a structured natural language error containing the bounding box $B_{conflict}$ back to the LLM for self-correction.}
\label{fig:pipeline}
\end{figure}

The core of our neuro-symbolic framework is an iterative loop that bridges the gap between neural generation and symbolic verification. 
The pipeline consists of five stages designed to guarantee physical safety. 
Its structure is illustrated in Figure \ref{fig:pipeline} and detailed in the protocol of Figure \ref{fig:workflow_protocol}.

\begin{figure}[htbp]
\centering
\begin{tcolorbox}[
    colback=white, 
    colframe=black!75!black, 
    title=\textbf{Protocol: The Evaluator-Refiner Workflow}, 
    fonttitle=\large\bfseries,
    arc=0mm, 
    outer arc=0mm,
    boxrule=0.8pt
]
\small
\begin{description}[style=multiline, leftmargin=3cm, font=\bfseries]
    \item[Step 1: Initialization] \textbf{Environment Grounding (The Generator) \cite{abdelaal2025gllm}} \\
    The user provides a natural language prompt defining the \textit{machining intent} and \textit{workspace topography}. The GLLM \cite{abdelaal2025gllm} synthesizes a G-code draft, while the system initializes the \textbf{Spatial Heap} by marking discrete coordinates as \texttt{Stock} or \texttt{Environment}.

    \item[Step 2: Discretization] \textbf{Kinematic Exhaustion (The Parser) \cite{lee2026cnc}} \\
    The continuous G-code is passed to the kinematic parser. To account for physical uncertainties, the parser applies a Minkowski sum dilation ($\epsilon$) to the tool geometry. The expanded continuous trajectory is then voxelized into a discrete 3D integer domain ($\mathbb{Z}^3$), converting geometric commands into formal spatial resource requests.

    \item[Step 3: Verification] \textbf{Formal Evaluation (The Evaluator) \cite{lee2026cnc}} \\
    The SL Prover verifies the discretized path step-by-step. Using the \textbf{Separating Conjunction} ($*$), it checks for disjointness between the requested tool volume ($V_{tool}$) and the environment ($V_{env}$), aiming to prove the state transition $\{P\}\mathbb{C}\{Q\}$.

    \item[Step 4: Feedback] \textbf{Self-Correction (The Refiner)} \\
    If the proof fails ($V_{tool} \cap V_{env} \neq \emptyset$), 
    a \textbf{Spatial Data Race} is triggered. The prover identifies the specific intersection and condenses the conflicting voxels into a minimal axis-aligned bounding box ($B_{conflict}$). This failure is translated into a structured error signal:
    \begin{center}
        \begin{minipage}{0.85\textwidth}
            \itshape ``Collision detected at line N45. \textbf{Spatial Data Race:} Tool intersects {\normalfont\texttt{Environment}} within 
            
            $B_{conflict} = [X_{45-55}, Y_{30-40}, Z_{0-10}]$.
            
            \textbf{Directive:} Regenerate path to bypass this restricted volume.''
        \end{minipage}
    \end{center}
    This deterministic signal is appended to the context window to guide targeted LLM refinement in the next iteration.

    \item[Step 5: Termination] \textbf{Safety Guarantee} \\
    The loop terminates only upon a successful formal proof of spatial disjointness across the entire program. The final output is a \textit{correct-by-construction} G-code sequence ready for physical execution.
\end{description}
\end{tcolorbox}
\caption{Detailed procedural steps of the Neuro-Symbolic Evaluator-Refiner pipeline, mapping natural language intent to mathematically verified machine instructions.}
\label{fig:workflow_protocol}
\end{figure}

\subsection{Physical Grounding via B-Rep Extraction}
\label{sec:brep_extraction}

A fundamental prerequisite for our neuro-symbolic architecture is establishing a mathematical ground truth of the physical machining environment. 
Our framework directly extracts the exact Boundary Representation (B-Rep) from standard 3D CAD models. We achieve this by utilizing OpenCASCADE \cite{opencascade}, an industrial-grade, open-source C++ geometric modeling kernel. OpenCASCADE provides a robust engine for interpreting complex ISO 10303-21 (STEP) text schemas, performing rigorous mathematical translations to build precise 3D topological structures in memory. 

Our extraction traverses the topological tree of the STEP file to isolate specific entities—such as Planes or Cylinders. By mapping these topological entities to their underlying mathematical definitions via the OpenCASCADE kernel, we extract geometric parameters. For instance, the pipeline retrieves the exact center coordinates $(x, y, z)$ and radii of cylindrical fixtures or stock boundaries. 

To validate this extraction methodology against industry standards, we utilize standard reference test models provided by the National Institute of Standards and Technology (NIST) \cite{nist_step_dataset}, specifically from the Communications Technology Laboratory, Smart Connected Systems Division, Smart Connected Manufacturing Systems Group.

By parsing these STEP files through our OpenCASCADE-driven pipeline, we isolate the exact coordinate boundaries of the target part. 
This  geometric data serves as the physical truth. 
It is subsequently discretized and padded via Minkowski sums to populate the Spatial Heap for the SL Prover, while simultaneously being formatted as structured text constraints to ground the LLM via Retrieval-Augmented Generation (RAG), as detailed in Section \ref{sec:rag_context}.

\subsubsection{Algorithmic Implementation of the Geometry Extractor}
\label{sec:extractor_logic}

To implement the extraction of the Boundary Representation (B-Rep), we implemented a parsing algorithm utilizing the OpenCASCADE geometric kernel. The core logic of this parser bridges the gap between the complex topological tree of a STEP file and the strict spatial constraints required by both our Large Language Model (LLM) and the Separation Logic (SL) prover. 

The algorithmic workflow proceeds through three primary phases:

\begin{enumerate}
    \item \textbf{Initialization and Root Transfer:} 
    The algorithm begins by ingesting the raw STEP file schema. The geometric kernel translates the textual STEP entities into a unified, mathematical 3D shape object in memory. This establishes the foundational topological hierarchy (the complete B-Rep) of the target part, unmachined stock, and environmental fixtures.

    \item \textbf{Topological Traversal:} 
    Because a physical CNC workspace consists of an assembly of various interacting geometric entities, the algorithm employs a topological explorer to systematically traverse the shape's hierarchical tree. By filtering the traversal specifically for face entities (e.g., \texttt{TopAbs\_FACE}), the algorithm isolates individual boundary surface of the 3D solid while ignoring internal construction curves or redundant vertices.

    \item \textbf{Geometric Adaptation and Parameter Extraction:} 
    The critical translation from topological boundaries to deterministic mathematics occurs in this final phase. For each isolated face, a surface adaptor evaluates the underlying continuous mathematical definition to classify the geometry type (e.g., analytical planes, cylinders, or spheres). Once a specific analytic surface is identified---such as a cylindrical clamp---the algorithm extracts its absolute spatial dimensions. For instance, it retrieves the exact $(x, y, z)$ center coordinates and the bounding radius. 
\end{enumerate}

The complete Python implementation executing this three-phase workflow via the \texttt{pythonocc-core} library is provided in Listing \ref{fig:python_extractor_code}.

\begin{lstlisting}[
    language=Python,
    caption={Python implementation of the B-Rep extraction algorithm mapping CAD topology to continuous spatial coordinates.},
    label={fig:python_extractor_code},
    basicstyle=\ttfamily\scriptsize,
    numbers=left,
    numberstyle=\tiny\color{gray},
    stepnumber=1,
    numbersep=8pt,
    frame=single,
    rulecolor=\color{black!30},
    breaklines=true,
    breakatwhitespace=true,
    keywordstyle=\color{blue}\bfseries,
    stringstyle=\color{purple},
    commentstyle=\color{green!50!black}\itshape,
    showstringspaces=false,
    columns=flexible,
    keepspaces=true
]
import sys
import os
from OCC.Core.STEPControl import STEPControl_Reader
from OCC.Core.TopExp import TopExp_Explorer
from OCC.Core.TopAbs import TopAbs_FACE, TopAbs_EDGE, TopAbs_VERTEX
from OCC.Core.BRepAdaptor import BRepAdaptor_Surface
from OCC.Core.GeomAbs import *

# Mapping OpenCASCADE numeric codes to human-readable text
SURFACE_TYPES = {
    GeomAbs_Plane: "PLANE (Flat Surface)",
    GeomAbs_Cylinder: "CYLINDER (Curved Tube)",
    GeomAbs_Cone: "CONE (Tapered Surface)",
    GeomAbs_Sphere: "SPHERE (Ball Surface)",
    GeomAbs_Torus: "TORUS (Donut/O-Ring Surface)",
    GeomAbs_BezierSurface: "BEZIER SURFACE (Smooth Freeform)",
    GeomAbs_BSplineSurface: "B-SPLINE / NURBS SURFACE (Complex Engineering Surface)",
    GeomAbs_SurfaceOfRevolution: "SURFACE OF REVOLUTION",
    GeomAbs_SurfaceOfExtrusion: "SURFACE OF EXTRUSION",
    GeomAbs_OtherSurface: "UNKNOWN COMPLEX SURFACE"
}

def log_output(message, file_handle):
    """Helper function to print to terminal AND write to the text file simultaneously."""
    print(message)
    file_handle.write(message + "\n")

def analyze_downloaded_step(file_path):
    if not os.path.exists(file_path):
        print("Error: The file '" + str(file_path) + "' does not exist.")
        return

    # Create the output text file path in the current working directory
    base_name = os.path.splitext(os.path.basename(file_path))[0]
    output_txt_path = os.path.join(os.getcwd(), base_name + "_report.txt")

    # Open the text file for writing
    with open(output_txt_path, "w", encoding="utf-8") as report_file:
        
        log_output("Loading STEP file: " + os.path.basename(file_path) + "...", report_file)
        reader = STEPControl_Reader()
        status = reader.ReadFile(file_path)
        
        if status != 1:
            log_output("Error: OpenCASCADE failed to parse this STEP file.", report_file)
            return
            
        reader.TransferRoots()
        shape = reader.OneShape()
        
        # 1. Total Summary Counters
        face_exp = TopExp_Explorer(shape, TopAbs_FACE)
        total_faces = sum(1 for _ in iter(face_exp.More, False) if not face_exp.Next())

        edge_exp = TopExp_Explorer(shape, TopAbs_EDGE)
        total_edges = sum(1 for _ in iter(edge_exp.More, False) if not edge_exp.Next())

        vtx_exp = TopExp_Explorer(shape, TopAbs_VERTEX)
        total_vtx = sum(1 for _ in iter(vtx_exp.More, False) if not vtx_exp.Next())
        
        log_output("\n========================================", report_file)
        log_output("       STEP FILE TOPOLOGY METRICS       ", report_file)
        log_output("========================================", report_file)
        log_output(" Total Unique Vertices (Points): " + str(total_vtx), report_file)
        log_output(" Total Unique Edges (Lines):     " + str(total_edges), report_file)
        log_output(" Total Unique Faces (Surfaces):  " + str(total_faces), report_file)
        log_output("========================================\n", report_file)

        # 2. Detailed Surface Breakdown (Processing ALL surfaces)
        log_output("--- SURFACE GEOMETRY DETECTED ---", report_file)
        face_exp.ReInit()
        face_idx = 1
        
        while face_exp.More(): 
            face = face_exp.Current()
            surf_adaptor = BRepAdaptor_Surface(face)
            surf_type = surf_adaptor.GetType()
            
            type_str = SURFACE_TYPES.get(surf_type, "Other/Complex")
            log_output(" Surface #{:03d}: {}".format(face_idx, type_str), report_file)
            
            # Extract Dimensions and Coordinates
            if surf_type == GeomAbs_Cylinder:
                cylinder = surf_adaptor.Cylinder()
                center = cylinder.Location()
                log_output("   -> Dimension: Radius = {:.2f} mm".format(cylinder.Radius()), report_file)
                log_output("   -> Center XYZ: ({:.3f}, {:.3f}, {:.3f})".format(center.X(), center.Y(), center.Z()), report_file)
                
            elif surf_type == GeomAbs_Sphere:
                sphere = surf_adaptor.Sphere()
                center = sphere.Location()
                log_output("   -> Dimension: Radius = {:.2f} mm".format(sphere.Radius()), report_file)
                log_output("   -> Center XYZ: ({:.3f}, {:.3f}, {:.3f})".format(center.X(), center.Y(), center.Z()), report_file)
                
            face_idx += 1
            face_exp.Next()
            
        log_output("\nAnalysis Complete.", report_file)
        print("\nSaved plain text report to: " + output_txt_path)

if __name__ == "__main__":
    if len(sys.argv) < 2:
        print("Error: Please provide the path to your STEP file.")
        print("Usage: python analyze_real_step.py /path/to/your/file.stp")
        sys.exit(1)
        
    target_file = sys.argv[1]
    analyze_downloaded_step(target_file)
\end{lstlisting}

These precise, continuous parameters constitute the deterministic workspace topography. This data (e.g., identifying a fixture as a cylinder located at $X=-245.0, Z=-100.0$ with $R=20.0$) is exported by the parser to be mathematically padded via Minkowski sums. The resulting bounds are then injected into the LLM's static context prompt and discretized to populate the SL Prover's Spatial Heap.

To demonstrate the efficacy of this pipeline, Figure \ref{fig:nist_extraction_output} shows a partial extraction report generated from a  STEP file \cite{nist_step_dataset}. 
As shown, the algorithm successfully bypasses abstract topology to expose the explicit center coordinates and dimensions of the part's physical features. 
These values are directly routed into the RAG context window and the Separation Logic prover.

\begin{figure}[htbp]
\centering
\begin{lstlisting}[
    basicstyle=\ttfamily\scriptsize,
    frame=single,
    rulecolor=\color{black!30},
    breaklines=true,
    columns=flexible,
    keepspaces=true,
    captionpos=b
]
Loading STEP file: nist_ctc_01_asme1_rd.stp...

========================================
       STEP FILE TOPOLOGY METRICS       
========================================
 Total Unique Vertices (Points): 1734
 Total Unique Edges (Lines):     856
 Total Unique Faces (Surfaces):  139
========================================

--- SURFACE GEOMETRY DETECTED ---
 Surface #001: PLANE (Flat Surface)
 Surface #002: PLANE (Flat Surface)
 Surface #003: PLANE (Flat Surface)
 Surface #004: PLANE (Flat Surface)
 Surface #005: PLANE (Flat Surface)
 Surface #006: CYLINDER (Curved Tube)
   -> Dimension: Radius = 20.00 mm
   -> Center XYZ: (-245.000, -0.000, -100.000)
 Surface #007: PLANE (Flat Surface)
 Surface #008: CYLINDER (Curved Tube)
   -> Dimension: Radius = 10.00 mm
   -> Center XYZ: (245.000, -15.000, -100.000)
 Surface #009: PLANE (Flat Surface)
 Surface #010: CYLINDER (Curved Tube)
   -> Dimension: Radius = 10.00 mm
   -> Center XYZ: (245.000, 15.000, -100.000)
 
 ... [Output truncated for brevity] ...

 Surface #137: PLANE (Flat Surface)
 Surface #138: CYLINDER (Curved Tube)
   -> Dimension: Radius = 4.89 mm
   -> Center XYZ: (398.000, -110.109, -18.007)
 Surface #139: PLANE (Flat Surface)

Analysis Complete.
\end{lstlisting}
\caption{Truncated extraction report for the NIST standard reference model (\texttt{nist\_ctc\_01\_asme1\_rd.stp}). The output demonstrates the successful translation of abstract CAD topology into explicit mathematical parameters.}
\label{fig:nist_extraction_output}
\end{figure}

\subsection{CAD-Augmented Generation (RAG)}
\label{sec:rag_context}

A fundamental limitation of applying Large Language Models (LLMs) to physical control systems is the hallucination problem. While LLMs excel at syntactic generation, they lack inherent spatial awareness. An unprompted LLM generating G-code frequently operates under the implicit assumption of an infinite, obstacle-free workspace. To ensure the initial generation is physically grounded and to minimize spatial hallucinations, our framework employs a specialized Retrieval-Augmented Generation (RAG) pipeline to construct a highly constrained system prompt derived directly from the physical ground truth extracted in Section \ref{sec:brep_extraction}.

\subsubsection{Topological Context Injection}
Before processing the user's machining intent, the system retrieves the explicit, deterministic continuous coordinates extracted from the CAD model and formats them into structured text within the LLM's context window. This static context comprises three primary domains:
\begin{itemize}
    \item \textbf{Kinematic Limits:} The absolute travel boundaries of the CNC machine bed, establishing the physical bounds for the spatial coordinate lattice.
    \item \textbf{Workspace Topography:} The coordinate boundaries defining the unmachined stock material and all workholding fixtures. Rather than relying on arbitrary estimates, these bounds are direct algorithmic translations of the analytical surfaces (e.g., the exact radii and center $(X,Y,Z)$ coordinates of cylindrical clamps) isolated by our STEP parser.
    \item \textbf{Tool Geometry:} The radius and functional length of the active cutting tool, which are critical for determining safe Z-axis retraction planes.
\end{itemize}

\subsubsection{Dynamic Margin Injection}
To closely align the neural generative model with our formal Separation Logic (SL) verifier \cite{lee2026cnc}, the RAG system dynamically adjusts the raw extracted CAD topography based on a required safety margin, denoted as $\epsilon$. In our formal SL model, physical machine uncertainties (e.g., servo lag, tool deflection) are mathematically mitigated by expanding spatial resources via a discrete Minkowski sum ($\oplus$) \cite{minarvcik2024minkowski}. 

To minimize the number of failure iterations required in the Evaluator-Refiner loop, we proactively translate these formal geometric expansions into LLM prompt constraints. For example, the system takes the raw deterministic coordinates extracted from the STEP file (such as the exact dimensions of Surface \#006 from our extraction) and applies a $B_{\epsilon}$ margin (e.g., $\epsilon = 2.0$ mm) to guarantee collision avoidance. 

These mathematically padded spatial dimensions are then injected directly into the prompt text. Consequently, the LLM optimizes its trajectory planning around these mathematically inflated physical obstacles, significantly increasing the probability of synthesizing a ``correct-by-construction'' toolpath on its very first pass.

\begin{figure}[htbp]
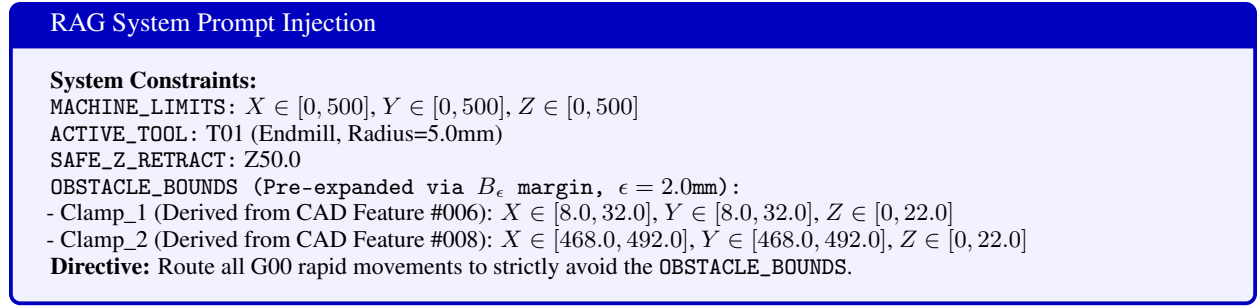

\centering
\begin{tcolorbox}[colback=blue!5!white,colframe=blue!75!black,title=RAG System Prompt Injection]
\small
\textbf{System Constraints:} \\
\texttt{MACHINE\_LIMITS:} $X \in [0, 500]$, $Y \in [0, 500]$, $Z \in [0, 500]$ \\
\texttt{ACTIVE\_TOOL:} T01 (Endmill, Radius=5.0mm) \\
\texttt{SAFE\_Z\_RETRACT:} Z50.0 \\
\texttt{OBSTACLE\_BOUNDS (Pre-expanded via $B_{\epsilon}$ margin, $\epsilon=2.0$mm):} \\
- Clamp\_1 (Derived from CAD Feature \#006): $X \in [8.0, 32.0]$, $Y \in [8.0, 32.0]$, $Z \in [0, 22.0]$ \\
- Clamp\_2 (Derived from CAD Feature \#008): $X \in [468.0, 492.0]$, $Y \in [468.0, 492.0]$, $Z \in [0, 22.0]$ \\
\textbf{Directive:} Route all G00 rapid movements to strictly avoid the \texttt{OBSTACLE\_BOUNDS}.
\end{tcolorbox}
\caption{Example of a dynamically injected RAG prompt. The injected bounds are derived from exact CAD extractions and padded via Minkowski sums to enforce mathematically rigorous physical constraints during code generation.}
    \label{fig:rag-prompt-injection}
\end{figure}

An example of the dynamically injected RAG prompt is illustrated in Figure \ref{fig:rag-prompt-injection}. Crucially, the constraints labeled \texttt{Clamp\_1} and \texttt{Clamp\_2} in this prompt are not arbitrary textual bounds; they are the direct algorithmic outputs of mapping the extracted STEP B-Rep topology into $\epsilon$-padded Cartesian bounding boxes.

\subsection{The Neuro-Symbolic Loop}
\label{sec:architecture_loop}

We enforce a strict architectural boundary to isolate the deterministic logic of the prover from the probabilistic generation of the Large Language Model. 
GLLM \cite{abdelaal2025gllm} and CNC controllers natively operate in a continuous floating-point domain (e.g., $\mathbb{R}^3$). Conversely, Separation Logic (SL) \cite{reynolds2002separation,o2004resources} is fundamentally designed to reason about discrete, countable resources. To reconcile this domain mismatch and enable automated self-correction, our framework implements a continuous neuro-symbolic iterative loop.

\subsubsection{The Forward Pass: Generation and Discretization}
The loop begins with the LLM generating a candidate sequence of G-code instructions based on the user's prompt. Before these instructions can be formally evaluated, they must be translated from continuous kinematics into discrete spatial logic. 

An intermediate Parser acts as this discretization layer. 
It calculates the intended trajectory and applies physical safety margins to encapsulate mechanical uncertainties. 
The Parser then maps this expanded continuous volume onto a discrete integer lattice ($\mathbb{Z}^{3}$) using the scaling multiplier function $\mathcal{S}_{grid}(x)$ \cite{lee2026cnc}. 
This process translates a continuous physical toolpath into a finite set of requested spatial coordinates, shielding the logic phase from floating-point complexities.

\subsubsection{The Backward Pass: Verification and Feedback}
Once discretized, the imperative G-code command is abstracted into a formal spatial resource request. The SL Prover evaluates this request statically. It uses the core operator of SL, the \textit{separating conjunction} ($*$), to assert that the requested discrete coordinates are strictly disjoint from any restricted environmental obstacles.

If the separating conjunction holds, the move is mathematically proven safe. However, if the requested coordinates intersect with pre-existing environmental obstacles, the separating conjunction fails. The Prover deterministically identifies this logical contradiction as a \textit{Spatial Data Race} \cite{lee2026cnc}. 

Rather than simply rejecting the code and prompting the LLM to "try again" blindly, the framework utilizes this logical failure to generate structured, actionable feedback. The system condenses the conflicting discrete coordinates into a localized bounding box ($B_{conflict}$) and automatically feeds this explicit spatial constraint back into the LLM's context window. This mathematically derived feedback guides the neural model to refine its trajectory, repeating the generative cycle until a formally verified, correct-by-construction toolpath is achieved. 

The precise mathematical semantics and inference rules governing this verification engine are detailed in the next section.

\section{The Symbolic Filter: Separation Logic for G-Code}

\subsection{The Spatial Heap}
\label{sec:spatial_heap}

Traditional Separation Logic (SL) \cite{reynolds2002separation,o2004resources} was engineered to reason about shared mutable data structures in software, where a logical ``heap'' maps discrete integer memory addresses to data values. 
Our former work executes a domain shift: we model the physical CNC workspace as a managed logical resource termed the \textit{Spatial Heap} ($h$) \cite{lee2026cnc}. 
By establishing a formal correspondence between physical occupancy and logical memory ownership, we enable the deterministic evaluation of physical safety.

Let the domain of discrete physical locations be defined by the integer grid $\mathbb{Z}^3$, generated by the Parser's discretization phase via the scaling multiplier function $\mathcal{S}_{grid}(x)$. Furthermore, the domain of valid physical object states is defined by the set:
\begin{equation}
    status = \{Tool, Environment, Stock, Empty\}
\end{equation} 
The spatial heap $h$ is formally modeled as a finite partial function mapping discrete physical coordinates to their specific occupancy status:
\begin{equation}
    h : \mathbb{Z}^3 \rightharpoonup status
\end{equation} 

Under this model, ownership of discrete spatial coordinates is strictly managed as a memory resource. To evaluate collision-free states, we rely on spatial independence. Two spatial heaps, $h_1$ and $h_2$, are considered mathematically disjoint, denoted as $h_1 \perp h_2$, if and only if their defined domains share no common coordinate addresses in $\mathbb{Z}^3$:
\begin{equation}
    h_1 \perp h_2 \iff dom(h_1) \cap dom(h_2) = \emptyset
\end{equation} 

When $h_1 \perp h_2$ holds, their disjoint union is denoted as $h_1 \uplus h_2$. This fundamental property of disjointness serves as the mathematical bedrock of our verification engine. By ensuring that the coordinate domain owned by the \texttt{Tool} is always disjoint from the domains owned by the \texttt{Environment} or unmachined \texttt{Stock}, the system can deterministically prove the absence of physical collisions prior to machine execution.

\subsection{Minkowski Sums for Physical Uncertainty}
\label{sec:minkowski}

A critical vulnerability in utilizing LLMs for physical control is their tendency to generate trajectories with zero-margin tolerances. An LLM may synthesize a mathematically valid toolpath that clears an obstacle by a fraction of a millimeter. While logically sound in a continuous domain, such paths are highly susceptible to physical collisions due to real-world mechanical uncertainties, including servo tracking error, spindle runout, and tool deflection. 

To bridge this gap between neural generation and physical realities, our framework statically integrates a worst-case safety margin. Rather than complicating the discrete logic of the Prover, we encapsulate this physical uncertainty geometrically during the Parser phase utilizing discrete Minkowski sums \cite{lee2026cnc}.

The discrete Minkowski sum of two sets of position vectors $A$ and $B$ in $\mathbb{Z}^3$ is defined as the addition of each vector in $A$ to each vector in $B$:
\begin{equation}
    A \oplus B = \{a + b \mid a \in A, b \in B\}
\end{equation}

To guarantee physical safety against mechanical deviations, we mathematically pad the spatial request. Let $V_{tool}$ represent the discrete geometric volume of the cutting tool, and $B_{\epsilon}$ represent a discrete axis-aligned bounding box using the $L_{\infty}$ Chebyshev norm of radius $\epsilon$, which defines the maximum allowable physical deviation:
\begin{equation}
    B_{\epsilon} \triangleq \{c \in \mathbb{Z}^3 \mid ||c||_{\infty} \le \epsilon\} = \{(x,y,z) \in \mathbb{Z}^3 \mid \max(|x|,|y|,|z|) \le \epsilon\}
\end{equation}

When the LLM commands a geometric path, the Parser first calculates the discrete base trajectory $path$ (e.g., $path_{lin}$ for linear interpolations or $path_{box}$ for rapid traversals) \cite{lee2026cnc}. The total required spatial footprint, $V_{path}$, is then generated by applying the discrete Minkowski sum:
\begin{equation}
    V_{path} = path \oplus V_{tool} \oplus B_{\epsilon}
\end{equation}

By expanding the spatial footprint, we effectively exhaust physical uncertainty into a bounded geometric envelope. Consequently, the SL Prover evaluates a padded spatial request. This architectural decision isolates complex continuous uncertainty from the discrete logical proof phase, allowing the Prover to operate strictly on coordinate sets while mathematically guaranteeing physical safety margins.

\subsection{The Parser-Prover Handshake}
\label{sec:parser_handshake}

We enforce a strict architectural boundary to isolate the discrete logic of the prover from continuous kinematic variables. Generative AI models and CNC controllers operate in a continuous floating-point domain (e.g., $\mathbb{R}^3$), generating coordinate strings such as \texttt{X12.345 Y50.000}. Conversely, Separation Logic is fundamentally designed to reason about discrete, countable resources. To reconcile this domain mismatch, we adopt the Parser-Prover Handshake introduced by Lee \cite{lee2026cnc}.

\subsubsection{Kinematic Discretization}
Before any G-code instruction reaches the SL Prover, the intermediate Parser extracts the LLM's generated coordinates and translates them into the finite coordinate set representing the swept volume $V_{swept}$ using the aforementioned $\mathcal{S}_{grid}(x)$ and Minkowski sum ($\oplus$). This ensures that the SL Prover evaluates purely discrete spatial sets, completely isolated from floating-point arithmetic complexities.

\subsubsection{Formal Evaluation via Separation Logic Triples}
Once discretized into a finite coordinate set (e.g., the discrete trajectory footprint $V_{path}$), the operation is formulated into standard Separation Logic Triples of the form $\{P\}\mathbb{C}\{Q\}$. In this neuro-symbolic handshake, the imperative G-code command $\mathbb{C}$ is abstracted into a discrete spatial resource request.

The SL Prover evaluates this request using the spatial resource allocation function $\mathcal{R}(C, \theta)$ and the basic points-to relation $c \mapsto \theta$ where $c \in C$ holds. 
The Prover operates by checking the following logical components:
\begin{itemize}
    \item \textbf{Precondition ($\{P\}$):} The assertion that the requested coordinate set $V_{path}$ must be completely disjoint from restricted environmental obstacles. This is mathematically verified by attempting to satisfy the separating conjunction ($*$) between the requested tool allocation and the known environment, such as asserting $\mathcal{R}(V_{start}, Tool) * \mathcal{R}(C_{env}, Environment)$.
    \item \textbf{Postcondition ($\{Q\}$):} If disjointness is proven, the state of the Spatial Heap ($h$) is updated to reflect that the tool safely traversed the space, modifying ownership states (e.g., transitioning \texttt{Stock} to \texttt{Empty} upon material removal) as it moves.
\end{itemize}

For example, when a rapid traverse command (\texttt{G00}) is generated, the Prover does not simulate the physical movement. Instead, it statically evaluates the disjointness of the pre-evaluated spatial footprint $V_{path} = path_{box} \oplus V_{tool} \oplus B_{\epsilon}$. If the requested spatial allocation intersects with the environment ($V_{path} \cap C_{env} \neq \emptyset$), the separating conjunction ($*$) fails. The Prover deterministically traps this logical contradiction as a Spatial Data Race \cite{lee2026cnc}, preventing physical execution and feeding the minimal conflict bounding box ($B_{conflict}$) back to the LLM for correction.

\subsection{G-Code as Separation Logic Triples}
\label{sec:gcode_triples}

With the physical workspace mathematically modeled as a Spatial Heap, we map imperative G-code commands generated by the LLM into formal state transitions. We utilize Separation Logic (SL) Hoare Triples \cite{hoare1969axiomatic,reynolds2002separation}, denoted as $\{P\}\mathbb{C}\{Q\}$, where $\mathbb{C}$ represents the command, $P$ is the precondition  and $Q$ is the postcondition after execution.

The primary advantage of employing Separation Logic is its native support for \textit{local reasoning} via the Separating Conjunction ($*$). When evaluating a motion command, the SL Prover does not need to assess the state of the entire CNC machine. It only requires proof that the specific spatial footprint requested by the tool is disjoint from the restricted environment. 

Let $c \mapsto \theta$ denote the logical assertion that a discrete coordinate $c \in \mathbb{Z}^3$ is entirely occupied by the specified physical state $\theta \in \{Tool, Environment, Stock, Empty\}$. We can formalize the fundamental G-code instructions as logical memory mutations utilizing the spatial resource allocation $\mathcal{R}(C, \theta)$.

\subsubsection{G00 Rapid Positioning (Safe Traversal)}
Because \texttt{G00} is non-cutting, the trajectory requires the path to be strictly free of obstacles. The rule requires the entire swept volume to be \texttt{Empty} and guarantees explicit deallocation of the tool's previous volume in the post-condition.

\begin{equation}
\frac{V_{path} \cap (C_{env} \cup C_{stock}) = \emptyset}{
\begin{gathered}
    \{ \mathcal{R}(V_{start}, Tool) * \mathcal{R}(V_{path} \setminus V_{start}, Empty) * \mathcal{R}(C_{env}, Environment) \} \\
    \text{G00}(V_{start}, V_{final}, V_{path}) \\
    \{ \mathcal{R}(V_{final}, Tool) * \mathcal{R}(V_{path} \setminus V_{final}, Empty) * \mathcal{R}(C_{env}, Environment) \}
\end{gathered}
}
\end{equation}

\begin{itemize}
    \item \textbf{Strict Disjointness:} Following the operational semantics of our prior work \cite{lee2026cnc}, any coordinate within $V_{path}$ that overlaps with $C_{env}$ or $C_{stock}$ triggers a deterministic Spatial Data Race. Because the Prover evaluates this without a variable store, this rule statically ensures traversal occurs only through verified free air.
\end{itemize}

\subsubsection{G01 Linear Interpolation (Cutting and Mutation)}
Because the continuous kinematic state has been completely abstracted away by the Parser, the logic models linear cutting simply as a triple-state mutation in the Spatial Heap: clearing the tool's trailing volume, transitioning traversed \texttt{Stock} to \texttt{Empty}, and allocating the new \texttt{Tool} position. This rule accommodates trajectories passing through both workpiece material and free air \cite{lee2026cnc}.

\begin{equation}
\frac{V_{path} \setminus V_{start} \subseteq C_{stock} \cup C_{empty}}{
\begin{gathered}
    \{ \mathcal{R}(V_{start}, Tool) * \mathcal{R}(C_{stock}, Stock) * \mathcal{R}(C_{empty}, Empty) * \mathcal{R}(C_{env}, Environment) \} \\
    \text{G01}(V_{start}, V_{final}, V_{path}) \\
    \{ \mathcal{R}(V_{final}, Tool) * \mathcal{R}(C_{stock} \setminus V_{path}, Stock) * \mathcal{R}(C_{empty} \cup (V_{path} \setminus V_{final}), Empty) * \mathcal{R}(C_{env}, Environment) \}
\end{gathered}
}
\end{equation}

\begin{itemize}
    \item \textbf{Deallocation and Consumption:} The swept volume $V_{path} \setminus V_{final}$ is transitioned to \texttt{Empty}, representing the physical removal of \texttt{Stock} and the movement of the tool to prevent trailing ``ghost tool'' artifacts.
    \item \textbf{Path Safety:} The precondition ensures the toolpath footprint remains disjoint from the \texttt{Environment}. Any intersection with static fixtures triggers the fault transition, rendering the separating conjunction ($*$) unsatisfiable.
\end{itemize}

\section{The Self-Corrective Feedback Loop}

\subsection{Detecting Spatial Data Races}
\label{sec:spatial_data_races}

In concurrent software verification, a data race occurs when two threads attempt to simultaneously access the same memory address without proper synchronization, violating mutual exclusion. We extend this paradigm to physical kinematics by formalizing a mechanical collision as a logical contradiction, termed a \textit{Spatial Data Race}. 

In our neuro-symbolic framework, a Spatial Data Race occurs when the generated toolpath attempts to claim spatial memory addresses (voxels) that are already strictly owned by an obstacle, such as a workholding fixture or the machine enclosure. 

Mathematically, this conflict is detected during the evaluation of the Separating Conjunction ($*$). Let $h_{env}$ represent the Spatial Heap containing all restricted environment structures, and let $h_{swept}$ represent the logical resource request generated by the tool's swept volume for a given G-code command. The SL Prover attempts to evaluate the disjoint union of these heaps:
\begin{equation}
    h_{swept} \uplus h_{env}
\end{equation}

By the axiomatic definition of Separation Logic, this conjunction holds true if and only if the domains of the two heaps are entirely disjoint. If the Large Language Model generates a hallucinated trajectory that intersects with a physical obstacle, the domains will overlap. We formalize this overlap as the conflict set $V_{conflict}$:
\begin{equation}
    V_{conflict} = dom(h_{swept}) \cap dom(h_{env})
\end{equation}

If $V_{conflict} \neq \emptyset$, the logical proof of safety deterministically fails. The system flags this logical contradiction as a Spatial Data Race. 
The SL Prover leverages the formal failure of the Separating Conjunction to isolate the exact coordinate boundaries of $V_{conflict}$. 

This precise mathematical localization of the error transforms a binary proof failure into a structured data vector. The isolated $\mathbb{Z}^3$ coordinates of the Spatial Data Race serve as the direct trigger and payload for the subsequent LLM refinement phase, ensuring that the generative model receives deterministic, localized feedback on exactly where its physical reasoning failed.

\subsection{Generating the Error Signal}
\label{sec:error_signal}

To close the neuro-symbolic feedback loop, the formal failure of the SL Prover must be mapped back into the generative model's native domain: natural language. 
LLMs process tokens sequentially and lack the inherent architecture to parse raw, high-dimensional logical states. Therefore, providing the LLM with a generic ``collision detected'' flag results in blind, stochastic regeneration. To enable directed self-correction, our system translates the isolated Spatial Data Race into a structured semantic prompt.

Upon the deterministic failure of the Separating Conjunction, the Prover isolates the conflict set $V_{conflict} \subset \mathbb{Z}^3$. Because $V_{conflict}$ may comprise thousands of individual voxels, injecting raw coordinate arrays into the LLM's context window would rapidly induce token exhaustion and degrade attention mechanisms. To optimize the prompt, the system extracts the minimal bounding box $B_{conflict}$ encompassing the Spatial Data Race:
\begin{equation}
\begin{split}
    B_{conflict} = [x_{min}, x_{max}] \times [y_{min}, y_{max}] \times [z_{min}, z_{max}] \\
    \text{where } \forall (x,y,z) \in V_{conflict}
\end{split}
\end{equation}

This dimensional reduction condenses complex collision geometry into a token-efficient spatial constraint. The system then populates a deterministic natural language template, generating a highly localized error signal. This structured feedback provides the LLM with three critical pieces of context: instruction localization, the precise boundaries of the logical contradiction, and an actionable refinement directive.

\begin{figure}[htbp]
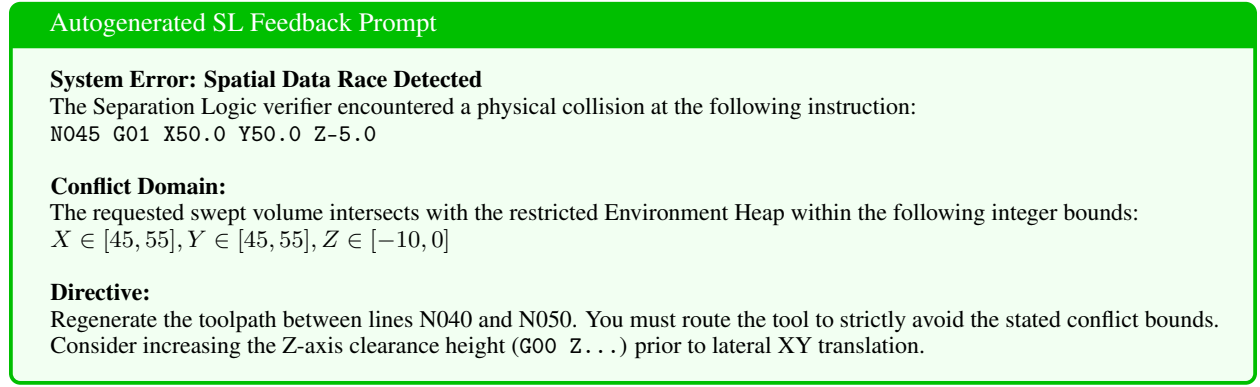

    \centering
\begin{tcolorbox}[colback=green!5!white,colframe=green!75!black,title=Autogenerated SL Feedback Prompt]
\small
\textbf{System Error: Spatial Data Race Detected} \\
The Separation Logic verifier encountered a physical collision at the following instruction: \\
\texttt{N045 G01 X50.0 Y50.0 Z-5.0} \\ \\
\textbf{Conflict Domain:} \\
The requested swept volume intersects with the restricted Environment Heap within the following integer bounds: \\
$X \in [45, 55], Y \in [45, 55], Z \in [-10, 0]$ \\ \\
\textbf{Directive:} \\
Regenerate the toolpath between lines N040 and N050. You must route the tool to strictly avoid the stated conflict bounds. Consider increasing the Z-axis clearance height (\texttt{G00 Z...}) prior to lateral XY translation.
\end{tcolorbox}
\caption{Self-corrective feedback loop output triggered by a physical collision (Logical Spatial Data Race) detected during symbolic execution.}
    \label{fig:sl-feedback-prompt}
\end{figure}

An example of the system-generated error signal, which is autonomously appended to the LLM's context window for the next iteration, is illustrated in Figure \ref{fig:sl-feedback-prompt}.

By transforming the mathematical failure of $h_{swept} \uplus h_{env}$ into this specific, bounds-aware natural language constraint, the framework restricts the LLM's probabilistic search space. This prevents the model from hallucinating repeated failures and forces convergence toward a mathematically verified, correct-by-construction toolpath.

\subsection{LLM Refinement and Self-Correction}
\label{sec:llm_refinement}

The final phase of the neuro-symbolic loop involves the ingestion of the Prover's error signal by the generative model to facilitate deterministic self-correction. When a Spatial Data Race is identified, the system does not simply discard the candidate toolpath. Instead, it constructs a cumulative refinement prompt consisting of the original user intent, the previously generated (and logically flawed) G-code sequence, and the structured error signal containing the conflict bounding box $B_{conflict}$. This aggregated data is appended directly to the LLM's context window.

By providing explicit mathematical bounds rather than vague failure warnings, the framework enables the LLM to engage in targeted in-context reflection. The model utilizes the restricted $\mathbb{Z}^3$ coordinates to recalculate its trajectory strategy. For instance, if the error signal indicates a logical collision within the lateral bounds of a toe clamp, the LLM can autonomously deduce the necessity of a vertical avoidance maneuver, seamlessly inserting a safe Z-axis retraction command (e.g., \texttt{G00 Z...}) prior to executing the XY translation. 

From an architectural standpoint, the injection of $B_{conflict}$ serves to drastically restrict the generative model's probabilistic search space. Unconstrained, an LLM might attempt to resolve a collision through stochastic trial-and-error, leading to high computational overhead and unpredictable convergence. However, the deterministic feedback provided by the SL Prover acts as a rigid semantic funnel. It penalizes the probability weights of generating coordinates within the restricted domain, heavily biasing the model toward mathematically viable, collision-free trajectories.

Once the LLM synthesizes the refined candidate code, the output is immediately routed back to the Parser for discretization and subsequently to the SL Prover for re-evaluation. This Evaluator-Refiner loop iterates continuously. Because the symbolic filter provides absolute, zero-false-positive spatial feedback, the LLM rapidly converges on a safe solution. The loop terminates strictly when the Separating Conjunction successfully joins the tool's swept volume with the environment heap, outputting a formal proof of safety alongside the deployable, correct-by-construction G-code.

\section{Conclusion}

This research establishes a transformative neuro-symbolic framework for verifiable G-code synthesis. By integrating the probabilistic generative capabilities of Large Language Models, such as those pioneered by the GLLM framework \cite{abdelaal2025gllm}, with a Separation Logic (SL) verifier for G-code \cite{lee2026cnc}, our architecture successfully yields formally verified, correct-by-construction toolpaths. This integration represents an advancement for autonomous manufacturing, bridging the semantic intent of neural generation with rigorous mathematical proofs of spatial disjointness.

To enable this neuro-symbolic evaluation, our framework builds upon a foundational domain shift that maps the physical CNC workspace to a logical \textit{Spatial Heap}, conceptualizing physical occupancy as logical memory ownership \cite{lee2026cnc}. Crucially, by evaluating physical collisions as formal \textit{Spatial Data Races} \cite{lee2026cnc}, the current work translates complex geometric intersections into binary logical contradictions. When the separating conjunction fails, these logical faults are condensed into minimal bounding boxes ($B_{conflict}$) and fed directly back to the LLM as structured spatial directives. This mechanism introduces an automated, deterministic self-correction loop that guarantees physical safety without relying on human-in-the-loop validation or the empirical path-similarity approximations utilized in purely neural approaches. 
Ultimately, this neuro-symbolic methodology offers a scalable, mathematically assured path toward fully autonomous, zero-collision CNC manufacturing.

\subsection{Future Work}
While this paper establishes the formal theoretical feasibility and the neuro-symbolic architecture of the Generator-Verifier loop, comprehensive experimental validation remains an ongoing effort. Future investigations will focus on transitioning this symbolic mapping into industrial applications across the following technical domains:

\begin{itemize}
    \item \textbf{Empirical Benchmarking and Voxelization Optimization:} Future work will quantify the real-world performance gains of our framework by benchmarking it against traditional CAM verification tools (e.g., VERICUT) across diverse, large-scale industrial G-code sequences. To mitigate potential state-space explosion as workspace volume and precision requirements increase, we will explore specialized solvers for quantized neural networks and optimized data structures for high-resolution voxel grids.
    
    \item \textbf{Concurrent Separation Logic for Multi-Agent Systems:} Modern CNC environments frequently utilize multiple independent kinematic chains, such as dual-spindle lathes or collaborative robotic cells. We plan to extend our results to Concurrent Separation Logic (CSL) \cite{o2004resources,brookes2016concurrent}. 
    
    \item \textbf{Intent Formalization:} To further close the ``intent gap'' between informal human instructions and machine-readable constraints, we will investigate advanced Retrieval-Augmented Generation (RAG) techniques. The goal is to automatically translate descriptions of complex machining tasks into structured Separation Logic specifications, further automating the safety-critical generation pipeline.
\end{itemize}

\bibliographystyle{unsrt}  
\bibliography{references}

\end{document}